\documentstyle[epsfig]{article}
\begin{document}
%
%
%
\def \vv#1{{\bf{#1}}}
\def \w#1{{\bf{#1}}}
\let\ov=\over
\let\lbar=\l
\let\l=\left
\let\r=\right
\def \der#1#2{{\partial{#1}\over\partial{#2}}}
\def \dder#1#2{{\partial^2{#1}\over\partial{#2}^2}}
\def \bb{\bigskip \goodbreak}
\def \R{{\rm I \! R}}
\def \E{{\cal E}}
\def \Oo{${\cal O}_0 \ $}
\def\be{\begin{equation}}
\def\ee{\end{equation}}

\def\encadre#1#2{\vbox{\hrule \hbox{ \kern-3pt \vrule  \kern3pt
 \vbox { \kern3pt
\vbox{\hsize #1 \noindent \strut #2 \strut }
   \kern3pt} \kern3pt \vrule} \hrule}}



\begin{titlepage}
\begin{center}
{\LARGE \bf SHORTCUT METHOD OF SOLUTION OF \\
GEODESIC EQUATIONS \\
\smallskip
\smallskip
FOR SCHWARZSCHILD BLACK HOLE}
\vfill
{\large Jean--Alain Marck} \\
\smallskip
D\'epartement d'Astrophysique Relativiste et de Cosmologie \\
 (UPR 176 du C.N.R.S.) \\
 Observatoire de Paris - Section de Meudon \\
 F-92195 Meudon Cedex, France \\
{\em e-mail} : {\em marck@obspm.fr}
\vfill
\vfill
To appear in {\em Classical and Quantum Gravity}
\vfill
\end{center}
\centerline{\bf Abstract}

\begin{quote}
It is shown how the use of the Kerr-Schild coordinate
system can greatly simplify the formulation of the geodesic equation of
the Schwarzschild solution. An application of this formulation to the
numerical computation of the aspect of a non-rotating black hole is
presented. The generalization to the case of the Kerr solution is presented
too.
\end{quote}

\begin{quote}
\end{quote}

\vfill
\end{titlepage}

\section{Introduction.} \label{intro}

\bigskip
Since the publication of the Schwarzschild static solution solution
\cite{sch16},
it has been well known that the geodesic equation in this space-time
can be solved analyticaly. The purpose of this note is to show that,
by the mean of the Eddington coordinate system, the equation of motion
of test particles in the Schwarzschild space-time can be greatly
simplified so as to provide a more efficient method of solution,
using ideas suggested by analysis of the less simple case of the
Kerr rotating black hole solution.

\medskip The forms of the spherical Schwarzschild metric as expressed
in term of the usual Schwarzschild coordinate system and in term of the
less usual Eddington \cite{edi24} coordinate system (which is
interpretable as a limiting case of the Kerr-Schild \cite{ks}
coordinate system for a rotating black hole) are presented in \S~2.
Taking advantage of the obvious constants of motion which appear in the
Schwarzschild space-time, we write down the geodesic equation in term
of the Eddington coordinate system as second order equations with
respect to some affine parameter.

\medskip
These results will be useful for many problems for which geodesic motion
is concerned. In particular, they can greatly simplify the numerical
integration of the test particles trajectory. As an application, we
present in \S~3 the apparent shape of a non-rotating Black Hole surrounded
by an accretion disc as seen by an observer who is travelling towards and
ultimately through the event horizon.

\section{The Schwarzschild metric and the geodesic equation.}

\bigskip
Using the standard Schwarzschild cordinate system
$( t , r , \theta , \varphi )$, the line element of the metric takes
the form

\be
  ds^2 = - (1 -{2M \over r} ) dt^2
	+ {1 \over (1 - \displaystyle{2M \over r}) } dr^2
	+ r^2 ( d\theta^2 + \sin^2 \theta d\varphi^2)
\ee
in units such that $c=G=1$, where $M$ is the mass of the Black
Hole. \par

\bigskip
\noindent The first integrals of the equations of motion are well known
to be expressible as

\begin{eqnarray}
  \dot t &= &{E \over (1 - {2M \over r} ) } \\ \nonumber
  r^4 \dot r^2 &=  &E^2 r^4 - (r^2 - 2Mr) (\mu^2 r^2 + K) \\ \nonumber
  r^4 \dot \theta^2 &= &K - {\Phi^2 \over \sin^2 \theta} \\ \nonumber
 \dot \varphi  &=  &{\Phi \over r^2 \sin^2 \theta}
\end{eqnarray}
where a dot denotes differenciation with respect to some affine parameter,
$\tau$ say, and where the constants of motion $\mu^2$, $E$ and $\Phi$ are
respectively the rest-mass, the energy and the angular momentum about the
axis $\sin\theta = 0$ of the test-particle, $K$ being the Carter's fourth
constant of motion \cite{car68} which reduces in this simple spherical
case to the square of the total angular momentum of the orbiting
particle. \par

\bigskip
Introducing the Eddington coordinate system $( T , x , y , z )$, {\sl i.e.}
the Kerr-Schild coordinate system for the limiting case of a non rotating
black hole, where $( x , y , z )$ are {\sl ``cartesian-like''} spatial
coordinates and where $T$ is a retarded time, which are related to the
Schwarzschild cordinates by means of

\be
  x = r \sin\theta \cos\varphi ; \quad
  y = r \sin\theta \sin\varphi ; \quad
  z = r \cos\theta
\ee
and
\be
  T = u-r \quad ; \quad du = dt
		+ {1 \over 1 - \displaystyle{2M \over r}} dr \ ,
\ee
the metric takes the (Kerr-Schild) form
\be
  ds^2 = - dT^2 + dx^2 + dy^2 + dz^2
	+ {2M \over r^3} \left( xdx + ydy + zdz + rdT \right)^2
\ee
where
\be
	r^2 = x^2 + y^2 + z^2 \ .
\ee
At this stage, it is to be noticed that the above coordinate transformation
(first published by Eddington in 1924 \cite{edi24}) solves the
{\sl ``Schwarzschild
singularity''} $(r=2M)$ problem in the sense of being regular accross the
event horizon $r = 2 M$.

\bigskip
In terms of this system, it is straightforward to show that the geodesic
equations reduce to \par

\bigskip
\centerline{
\encadre{7cm}{
\begin{eqnarray} \label{geo1}
  \dot T &= &{2 M K \over r^3 (E - \dot r)} + E \\
\label{geo2}
  \ddot x^i  &= &-3 K M {x^i \over r^5} - \mu^2 M {x^i \over r^3}
\end{eqnarray}
}
}
\bigskip
\noindent where $x^i := (x,y,z)$.

\bigskip
Keeping in mind that $\vert \dot r \vert = E$ when any test-particle
reaches the event horizon, the above equations show clearly the
irreversible nature of the Schwarzschild Black Hole. Moreover, unlike what
is obtained in terms of the Schwarzschild cordinates system, for which one
has to take care of the axis $\sin\theta = 0$, of the sign of $\dot r$
and $\dot \theta$ and of the presence of the event horizon, such
differential equations can be numericaly solved in a straightforward way
in order to determine the path of any test-particle.

\section{Apparent shape of a non-rotating Black Hole.}

\bigskip
Accretion disc are currently supposed to play an important role in
several astrophysical situations, especially when high-energy phenomena
are involved. Because of the astrophysical interest of such objects,
several authors have computed the apparent shape of of black hole with
thin accretion disc as seen from infinity (\cite{bar73} \cite{cun73}
\cite{lum79}). The main
purpose of these calculation was to simulate line profiles which are
commonly observed from astrophysical sources which are generally
interpreted as emission from an accretion disc around a
compact object which may be a black hole. Our present purpose is quite
different in the sense that, as an application of the previous
formul\ae, we will present the apparent shape of a black hole
surrounded by a thin accretion disc as seen by an observer who
is flying near the hole. \par

We will assume that the disc is a stationnary Keplerian one, orbiting in
the plane $z=0$. Each particle of the disc then follows a circular orbit
with angular velocity

\be
	\Omega := {d\varphi \over dt} = \sqrt{ {M \over r^3} } \ .
\ee

\bigskip
Following Page and Thorne \cite{pag74}, the flux of radiation from the surface
of the disc is given by

\be \label{f_e}
 F_e = {3 M \dot M \over 8\pi} {1 \over (\rho^2 - 3) \rho^5}
	\left\{ \rho - \sqrt{6}
	  + {\sqrt{3} \over 2} \log
		\left( (3-2\sqrt{2}) {\rho + \sqrt{3}
					\over \rho - \sqrt{3}}
		\right)
	\right\}
\ee
where $\dot M$ is the accretion rate and where we have introduced

\be
	\rho = \sqrt{r \over M} \ .
\ee
As shown by Ellis \cite{ell71}, the observed bolometric flux $F_o$ is given by

\be \label{r1}
	F_o = {F_e \over (1+z)^4}
\ee
where the redshift factor $(1+z)$ is given by the ratio of the two scalar
products

\be \label{r2}
	(1+z) = { {p_\alpha w^\alpha } \vert_{emission}
		\over {p_\alpha u^\alpha} \vert_{reception} }	\ ,
\ee
{\bf p}, {\bf w} and {\bf u} being the four-velocities of the photon, the
emitting particle and the observer respectively. This redshift factor
consists of a gravitational part due to the gravitational field of the
black hole (which is measurable only in the closed vicinity of the hole),
a Doppler part due to the rotation of the disc (the dominant one) and a
Doppler part due to the motion of the observer.

\begin{figure}
\epsfysize=10cm			
\centerline{\epsfbox{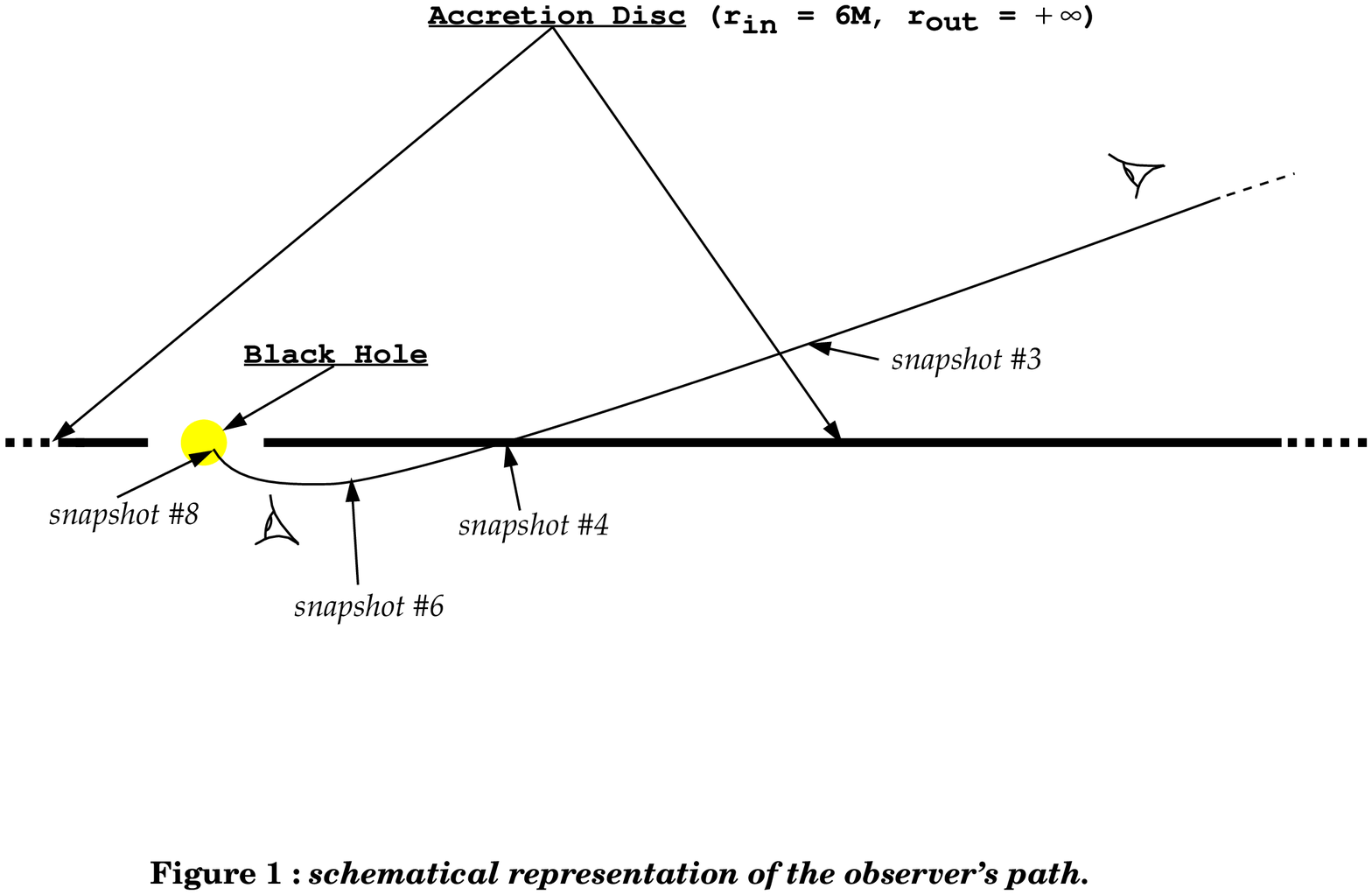}}	
\end{figure}

\bigskip
It is now straightforward to compute how the black hole will look.
Assuming that the observer takes a camera with him, the coordinates
of each pixel of the plate of the camera allows to compute the components
of the four-velocity of the photon that reaches that pixel (recall that
the four-velocity of a photon has only two true degrees of freedom). The
four velocity of each photon reaching the eyes of the observer can be
obtained in terms of an orthonormal parallel-propagated frame
$( \underline\lambda_0\ ,\  \underline\lambda_1\ ,\ \underline\lambda_2\ ,\
 \underline\lambda_3 )$ along the observer's trajectory as
\be
p^\mu = \lambda_0^\mu
	+ \cos\vartheta \lambda_2^\mu
	+ \sin\vartheta \cos\varphi \lambda_1^\mu
	+ \sin\vartheta \sin\varphi \lambda_3^\mu \ ,
\ee
where $\vartheta$ and $\varphi$ are two angles describing the photographic
plate in the rest frame of the observer. \par

A direct numerical integration of equations (\ref{geo1}) (\ref{geo2})
in the
special case $\mu^2 = 0$ for negative values of the affine parameter
$\tau$ gives the history of the photon and then the luminosity of the
source of its origin.  Finally, application of the correction factor
$(1+z)^4$ as given by equations (\ref{r1}) and (\ref{r2}) gives
 the brightness of this fiducial pixel. \par

\bigskip
The figures represent eight simulated photographs obtained by this procedure.
These ``snapshots'' have been computed at succesive steps
during the flight of an observer who is on free fall along a parabolic
orbit ($\mu^2=1\ ,\ E=1,\ K=12)$ of the Schwarzschild space-time (see figure
1).
On the first
snapshop, the observer is above the disc at a distant $r=1280 M$. The
observer then crosses the disc (snapshot \# 4, $r=39 M$) and, finally,
goes into the black hole (snapshot \#8, $r = 0.7 M$).
During this trip, the observer directs his camera in the direction of his
motion except on the last snapshot for which the observer directs his
camera toward the exterior of the hole. \par

The brightness of the pictures has been computed by means of
formul{\ae} (\ref{f_e}) and (\ref{r1}) while the coloration has been
added artificially.  The apparent position of distant stars has
also been computed. This helps to show the gravitational lense effect
which is more conspicuous on snapshot \# 2 ($r = 121 M$): some distant stars,
which are more or less perfectly aligned with the observer--black hole
system, appear as pieces of rings in the sky. The opacity
of the disc has been extremely minimized in order to make more visible
the second (and even third) image of the disc and the distant stars.
Moreover, this allows to show both the apparence of the upper and down
sides of the disc.
\par

\begin{figure}
\epsfysize=10cm			
\centerline{\epsfbox{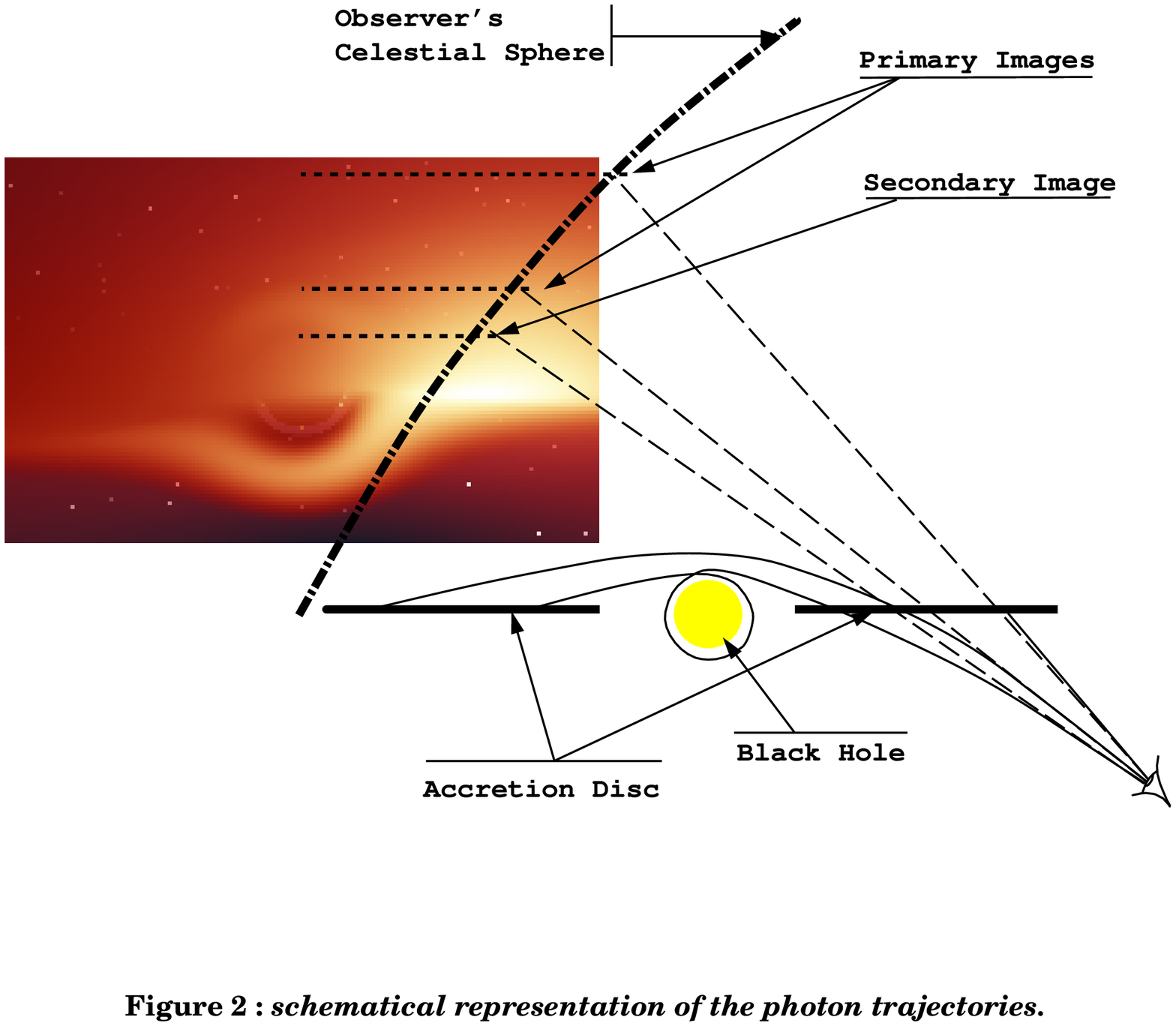}}	
\end{figure}

The trajectories of the photons and the link between these trajectories
and the apparent shape of the hole as seen by the observer are
schematically given in figure 2. Examination of figure 2 helps to understand
the apparent shape of the disc. Some of the photons emitted by the disc are
strongly deflected by the gravitational field (not the curvature) before
reaching the eyes of the observer. Hence, the observer is able to see the
upper rear side of the disc which appears like surrouding the top of the
black hole
({\sl primary image} on figure 2). In the same way, the observer can see
the down rear side of the disc which seems to surround the bottom of the
black hole. Now, some photons whose impact parameter is very close to the
capturing one make one (even two) turn(s) around the black hole before
reaching the eyes of the observer. This leads to the formation of a second
(third) image of the disc.  \par

\bigskip
The numerical scheme used in these calculation is of second order. The
integration has been done where respect to the parameter $\lambda$ which
is related to the affine parameter by
\be
	d\lambda = {d\tau \over r^2}
\ee
and which is well suited to the form of equations (\ref{geo1}--\ref{geo2})
in the sense that
it exploits the fact that, far from the hole, null
geodesics are straight lines. The code is fully parallelized. Vectorisation
is also possible but less efficient since the number and the kind of
operations needed to compute the value of each pixel of the screen vary
from a pixel to another one. \par

\section{Conclusion.}

\bigskip
We have shown that the use of the Eddington coordinates system in the
Schwarzschild space-time facilitates the description of the
geodesical motion in that space-time. This formulation is well adapted
to direct numerical integration because it avoid the usual
troubles tied to the {\sl spherical--type} coordinates system and the
trouble tied to the {\sl pseudo--singularity} $r = 2M$. \par

Such a formulation can be extended to the non-spherical case of the Kerr
black hole solution.
One can show that, in the Kerr-Schild coordinate system, the equations
of motion of a test particle of zero rest-mass can be written in the form:

\begin{eqnarray}
\Sigma^3 {\ddot x \over M} &= 		
	&- 4ar {Q\over\Delta} \Sigma \dot y \\ \nonumber
	& &+ (\Sigma - 4r^2) \sin\theta\cos\psi
	   \left\{K-\left(a{Q\over\Delta}\right)^2\right\} \\ \nonumber
	& &- ar\sin\theta\sin\psi {Q\over\Delta}
	   \left\{4(E\Sigma-Q) + (4a^2-\Sigma){Q\over\Delta}\right\} \\
\Sigma^3 {\ddot y \over M} &= 		
	&+ 4ar {Q\over\Delta} \Sigma \dot x \\ \nonumber
	& &+ (\Sigma - 4r^2) \sin\theta\sin\psi
	   \left\{K-\left(a{Q\over\Delta}\right)^2\right\} \\ \nonumber
	& &+ ar\sin\theta\cos\psi {Q\over\Delta}
	   \left\{4(E\Sigma-Q) + (4a^2-\Sigma){Q\over\Delta}\right\} \\
\Sigma^3 {\ddot z \over M} &= 		
	& - K \cos\theta (3r^2 -a^2\cos^2\theta) \\
\dot T &= & { 2 M K r \over \Sigma ( {\cal E} - \Sigma \dot r)}
		+ E 		
\end{eqnarray}
where we have introduced the quantities

\begin{eqnarray}
	\Sigma &= &r^2 + a^2\cos^2\theta \ , \\
	{\cal E} &= &E(r^2 + a^2)^2 - a\Phi \ , \\
	Q &= &\Sigma \dot r + {\cal E}
\end{eqnarray}
These
results have been used in the study of profiles and shifts of lines
emitted from Keplerian accretion disc around in X-ray binaries containing
a neutron star or a black hole \cite{ham94} and to the study of microlensing
effects in active galactic nuclei \cite{jar94}. \par

\begin{figure}
\epsfysize=20cm			
\end{figure}

\begin{figure}
\epsfysize=20cm			
\end{figure}

\begin{figure}
\epsfysize=20cm			
\end{figure}

\begin{figure}
\epsfysize=20cm			
\end{figure}

\bigskip\bigskip

\centerline{\large \bf Acknowledgments}

\bigskip
I wish to thank B. Carter, for helpful discussions. I am also
grateful to {\sl Silicon Graphics France} for the loan of computers. \par

\end{document}